\documentstyle[prb,twocolumn,epsfig,amssymb,floats,aps]{revtex}
\begin{document}
\draft

\twocolumn[\hsize\textwidth\columnwidth\hsize\csname @twocolumnfalse\endcsname

\title{High-energy magnon dispersion in the half-filled Hubbard model:
A comparison with La$_2$CuO$_4$}

\author{Pinaki Sengupta, Richard T. Scalettar and Rajiv R. P. Singh}
 
\address{Department of Physics, University of California, Davis, CA 95616}

\date{\today}

\maketitle

\begin{abstract}
We use quantum Monte Carlo methods and single-mode approximation
to study the magnon dispersion 
in the 2D half-filled Hubbard and phonon-coupled Heisenberg models.
We find that in the Hubbard model with $U/t< 8$, high-energy magnon
dispersion is similar to those observed in inelastic neutron scattering experiments in
$\mbox{La}_2\mbox{CuO}_4$. On the other hand, our studies of a
2D Heisenberg model coupled to dynamic optical bond phonons, fails to
reproduce the experimental dispersion. These results can be interpreted
as evidence for intermediate $U/t$ and charge fluctuations in the cuprate
materials.
\end{abstract}

\vskip2mm]

While there is still no consensus on the microscopic origin of 
superconductivity in high-temperature copper oxide superconductors,
it is widely believed that magnetic fluctuations play an important role.
This has led to extensive studies of magnetic fluctuations in the
parent compounds of these materials, such as $\mbox{La}_2\mbox{CuO}_4$.
The low-energy long-wavelength properties of these layered
compounds are well described by the non-linear sigma model and
the renormalized classical theory \cite{CHN}.
However, short wavelength or high energy spin-fluctuations may also be
relevant to superconductivity, which only sets in after long-range
antiferromagnetic order is lost. The description of these excitations
has raised many questions. Are there well-defined magnons at short
wavelengths? Is there significant spectral weight in multimagnon excitations?
Do the spinons present a better description of short-wavelength excitations?
Is there a coexistence of spinons and magnons? The most quantitative theoretical studies
have been done for the nearest-neighbor Heisenberg model, where one finds
magnons throughout the Brillouin Zone, with significantly reduced spectral
weight along the antiferromagnetic zone boundary \cite{series,hqmc}.

On the experimental side, Raman scattering has long provided evidence for
anomalous short-wavelength spin fluctuations \cite{SFLS}. The line-shape of the two-magnon
Raman spectra has not been adequately explained. Similar results are inferred
from optical absorption spectra \cite{gruninger}. Understanding these spectra is complicated
by the fact that light couples directly to the charge degrees of freedom and
thus their understanding within a spin-only picture requires many assumptions \cite{shraiman}.
In contrast, neutron scattering provides a direct probe of the
spin excitations \cite{swexpt1,swexpt2}. Recently,
the high energy magnon dispersion has been measured in La$_2$CuO$_4$ by inelastic neutron
scattering. One finds magnon dispersion at short wavelengths, with ($\pi/2,\pi/2$)
magnon energy about $13\%$ lower than that at ($\pi,0$) \cite{la2cuo4}. This is in contrast to
results for the nearest-neighbor Heisenberg model, where one finds the magnon energy
at ($\pi,0$) to be $7-10\%$ lower than at ($\pi/2,\pi/2$) \cite{series,hqmc}. 
The latter results have
also been seen experimentally in the materials
$\mbox{Cu(DCOO)}_2.4\mbox{D}_2\mbox{O}$ \cite{cftd} and 
$\mbox{Sr}_2\mbox{Cu}_3\mbox{O}_4\mbox{Cl}_2$ \cite{srcuocl}.

Explanation for the anomalous dispersion in
$\mbox{La}_2\mbox{CuO}_4$ has evoked considerable interest. Adding
a second neighbor antiferromagnetic exchange takes the spectra in the
opposite direction, ruling out the simplest possibility. One mechanism
that has been suggested is ring exchange of four electrons in a plaquette \cite{la2cuo4}.
Such a term can be directly obtained from the Hubbard model\cite{girvin}
and was first proposed for the high-$T_c$ materials to explain the Raman
spectra more than a decade ago \cite{ring1}. One interesting aspect of the ring-exchange
term is that, when treated fully, it does not change the spin-wave dispersion within
linear spin-wave theory.
Thus its effect on the dispersion is a purely quantum effect.
Recent numerical studies have focused on whether
such a term can explain the experimental dispersion without destabilizing
N\'{e}el order \cite{ring2}.

In a recent paper, Peres and Ara\'{u}jo\cite{peres} have studied the
2D Hubbard model using mean-field theory and obtained a 
dispersion relation similar to that observed in $\mbox{La}_2\mbox{CuO}_4$ 
for an intermediate value of the on-site 
interaction parameter, $U/t=6$, indicating that charge fluctuations -- ignored in 
the Heisenberg model -- need to be taken into account to explain the 
properties of this material.
Given the inherently approximate nature of the mean-field theory, it is 
important to confirm and extend this result using independent approaches.
We have used the determinant quantum Monte Carlo (det QMC) method 
and single mode approximation (SMA) to study the magnon dispersion in the
2D Hubbard model at half-filling. The det QMC method has been extensively used
to study the ground state properties of the 2D Hubbard model at half-filling
where the fermion sign problem can be avoided. We have studied
the magnon dispersion along the magnetic
zone boundary as a function of the on-site interaction parameter, $U/t$.
We find that with decreasing $U$, the zone-boundary dispersion changes sign
and for $U/t<8$ it becomes similar to those observed in $\mbox{La}_2\mbox{CuO}_4$.

Coupling to phonons is another potential source of anomalous dispersion in
the cuprates, because of the rather high magnon energies. We have used
the stochastic series expansion QMC to study a Heisenberg model coupled 
to optical bond phonons. We find that such a model fails to reproduce the
experimental spectra. While suggestive, this, however, does not rule out the possibility that
a more realistic treatment of the spin-phonon couplings can mimic the
experimental results.

The Hubbard model in two dimensions is given by the Hamiltonian
\begin{eqnarray}
H & = & -t\sum_{\langle{\bf i,j}\rangle\sigma}(c^{\dagger}_{{\bf i},\sigma}c_{\bf{j},\sigma}  
+ c^{\dagger}_{{\bf j},\sigma}c_{{\bf i},\sigma})\nonumber\\
  &   & +U\sum_{{\bf i}}(n_{{\bf i},\uparrow}-\frac{1}{2})(n_{{\bf i},\downarrow}-\frac{1}{2}) 
+ \mu\sum_{{\bf i}}n_{{\bf i}},
\label{H}
\end{eqnarray}
where $c^{\dagger}_{{\bf i},\sigma}(c_{{\bf i},\sigma})$ creates (annihilates) an 
electron with spin $\sigma$ at lattice site ${\bf i}$. The kinetic energy term
includes a sum over nearest neighbors $\langle{\bf i,j}\rangle$ and $t$ is the 
hopping integral between adjacent sites. $U$ is the on-site interaction, and 
$\mu$ is the chemical potential. We shall be dealing solely with the half-filled
band, ie., $\langle n_{{\bf i},\uparrow}+ n_{{\bf i},\downarrow} \rangle =1$. With 
the interaction term written in a particle-hole symmetric from as above, the half-filled
band corresponds to setting $\mu=0$. Henceforth we set $t=1$ and express
the interaction parameter $U$ in units of $t$.

In the limit of large $U$, the Hubbard model at half-filling maps on to the
Heisenberg model with exchange parameter $J=4t^2/U$. The Heisenberg model in
2D is given by the Hamiltonian
\begin{equation}
H = J\sum_{\bf i,j}{\bf S_i.S_j}
\end{equation}
The spin operators are related to the electron creation and annihilation
operators by the relation 
\[
{\bf S_i} = {1\over 2}c_{{\bf i},\sigma}^{\dagger}{\vec{\tau}}_{\sigma,\sigma'}c_{{\bf i},\sigma}
\]
where ${\vec{\tau}}$ are the Pauli matrices.

\begin{figure}
\centering
\epsfxsize=8.3cm
\leavevmode
\epsffile{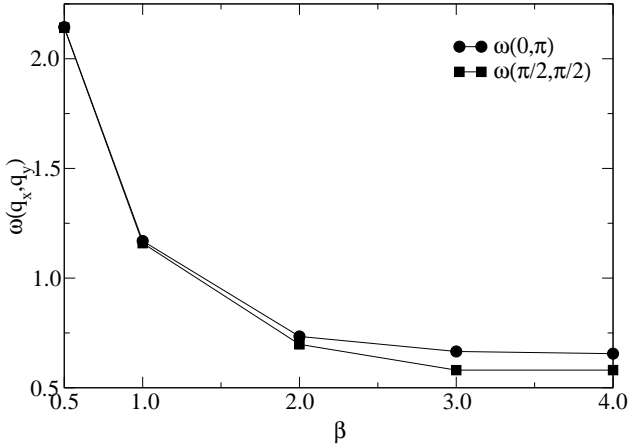}
\vskip1mm
\caption{Convergence of magnon energy to the ground state value as a function 
of the inverse temperature for a lattice of size $N$=8x8 at $U$=6.0. 
Error bars are smaller than symbol sizes.} 
\label{fig:fig1}
\end{figure}

The det QMC\cite{dqmc} used to study the 2D Hubbard model  is a finite temperature 
method that is based on discretizing
the imaginary time $\beta=L\Delta\tau$ and employing Trotter approximation to
decompose the full imaginary-time evolution. This approach treats the electron-electron
interactions exactly and at half-filling, where we focus in this work, is able to
produce results with small statistical fluctuations. Ground state expectation
values are obtained using sufficiently large values of $\beta$. Like most other
numerical approaches, the technique is limited to finite-size lattices and we 
have been able to study primarily two different lattice sizes, $N$=6x6 and $N$=8x8.
While this is not sufficient to do a complete finite-size scaling analysis to 
obtain thermodynamic expectation values, previous studies have shown that a lattice
of size $N$=8x8 is large enough to give reasonably good estimates of the thermodynamic
limit, especially when measuring quantities away from the antiferromagnetic
wave vector.

The Heisenberg model has been studied using the Stochastic Series Expansion (SSE)
QMC method\cite{sse1}. The SSE method is also a finite-temperature QMC method based on the 
importance sampling of the diagonal matrix elements of the Taylor expansion 
of $e^{-\beta H}$. Ground state expectation values can be obtained by using sufficiently
large values of $\beta$, and there are no approximations beyond statistical 
errors. With the recently developed ``operator loop update''\cite{sseloop}, the method has
proven to be very efficient tool for studying several different models.

In both approaches, we have measured the equal-time spin-spin correlation function 
$\langle S_{\bf i}S_{\bf j}\rangle$ and the associated static spin
structure factor in the momentum space given by
\begin{equation}
S({\bf q}) = {1\over N}\sum_{\bf i,j}e^{\bf q.(i-j)}\langle(n_{{\bf i}\uparrow}
-n_{{\bf i}\downarrow})(n_{{\bf j}\uparrow}-n_{{\bf j}\downarrow})\rangle
\end{equation}

\noindent We also evaluate the static spin susceptibility in the momentum space given by the
Kubo integral
\begin{equation}
\chi({\bf q}) = \int_0^{\beta} d\tau \langle S(-{\bf q},\tau)S({\bf q},0)\rangle
\end{equation}
The magnon energy is calculated using the relation \cite{sudbo}
\begin{equation}
\omega({\bf q}) = 2S({\bf q})/\chi({\bf q}).
\end{equation}

\begin{figure}
\centering
\epsfxsize=8.8cm
\leavevmode
\epsffile{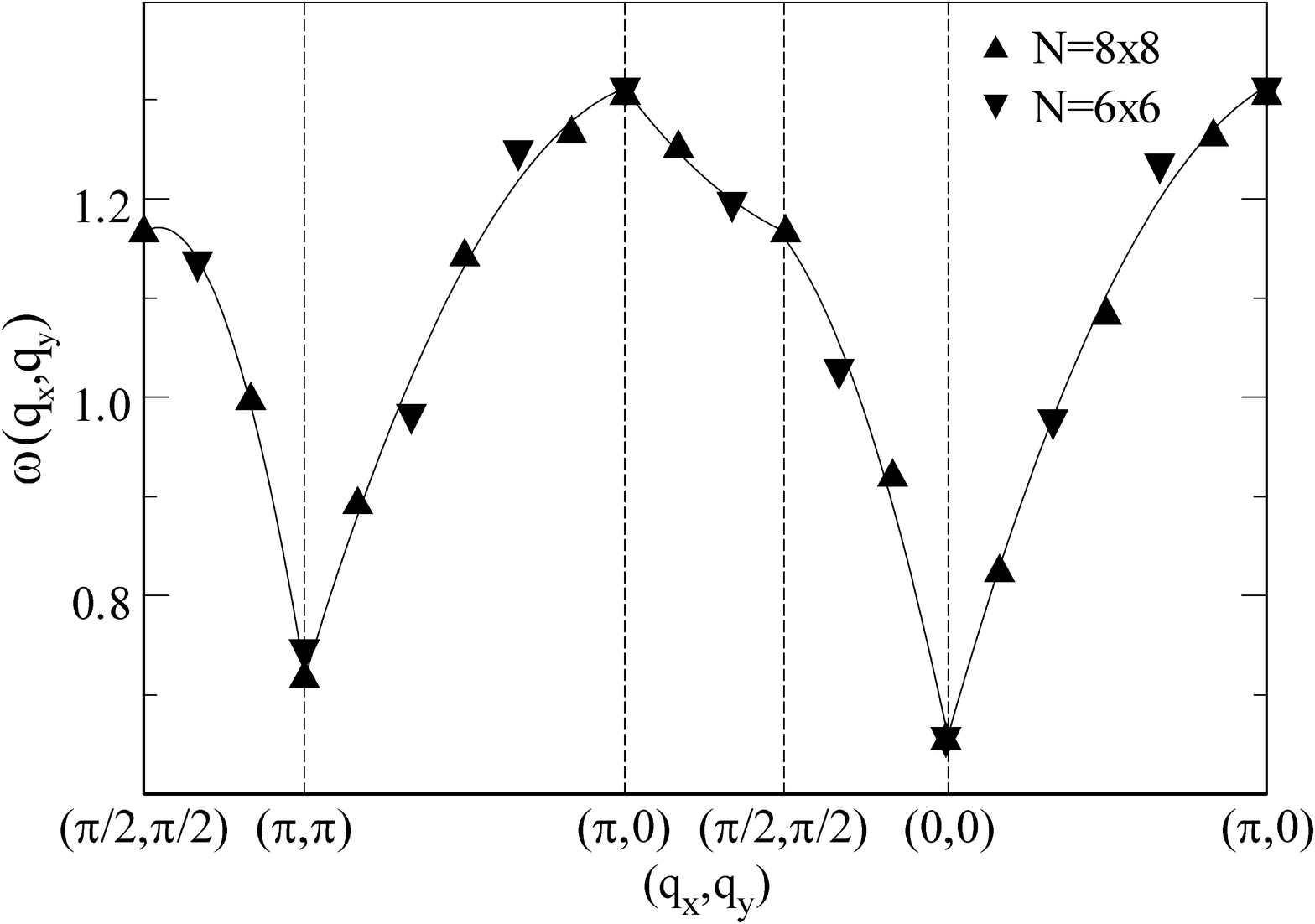}
\vskip-1mm
\caption{The magnon dispersion along a path similar to Figure 3a in ref. 9 for 6x6 and 8x8 lattices at $U=6.0$ and $\beta=3.0$. Error bars are smaller than
symbol sizes.} 
\label{fig:fig2}
\end{figure}

We start the discussion of the numerical results by studying the convergence of the
calculated quantities as a function of the inverse temperature, $\beta$. Figure \ref{fig:fig1}
shows a plot of the magnon energies at the edge, $(0,\pi)$, and the center,
$(\pi/2,\pi/2)$, of the zone boundary as a function of $\beta$ for $U=6.0$. 
The magnon energies are seen to converge to their $T=0$ values 
fairly rapidly with $\beta$. This convergence at relatively 
small values of $\beta$ is not surprising given that we only deal with 
high-energy quantities. Henceforth, all the
results presented are for $\beta=3.0$.

Next we present the results for the magnon energies along the entire magnetic zone
boundary for $U=6.0$. This value of $U$ is consistent with estimates of the effective on-site
energy in $\mbox{La}_2\mbox{CuO}_4$ obtained from photoemission\cite{photo} and optical\cite{optical}
spectroscopy data. Figure \ref{fig:fig2} shows the variation of the magnon energy for a path along the
Brillouin zone similar to the one considered in Figure 3a of ref.\onlinecite{la2cuo4} for two different
lattice sizes. Note that the data near ($\pi,\pi$) and ($0,0$) are strongly
temperature dependent, the latter being exactly proportional to temperature.
However, as shown in Figure \ref{fig:fig1}, the data along the zone boundary from ($\pi,0$) to
($\pi/2,\pi/2$) have reached close to their zero temperature values. The qualitative nature 
of the variation of the magnon energy along the path is in good agreement with the
experimental results and mean-field predictions. In particular, the magnon energy
at the center of the zone boundary, $(\pi/2,\pi/2)$, is found to be lower 
than that at the edge, $(0,\pi)$, as seen in the experimental
data and opposite to the dispersion found for the Heisenberg model. The magnitude
of the deviation is estimated to be $\approx -12\%$ of the $N$=8x8 system, 
matching closely the experimental observation. While this quantitative agreement with
experimental data is not conclusive-- the magnon energies were 
evaluated in our calculation using the SMA, and the finite-size effects 
are fairly large -- it is safe to conclude that the 2D Hubbard model,
with a value of the on-site interaction energy in the range estimated from photoemission and
optical spectroscopy data, can reproduce the correct magnon 
dispersion along the magnetic zone boundary.

\begin{figure}
\centering
\epsfxsize=8.3cm
\leavevmode
\epsffile{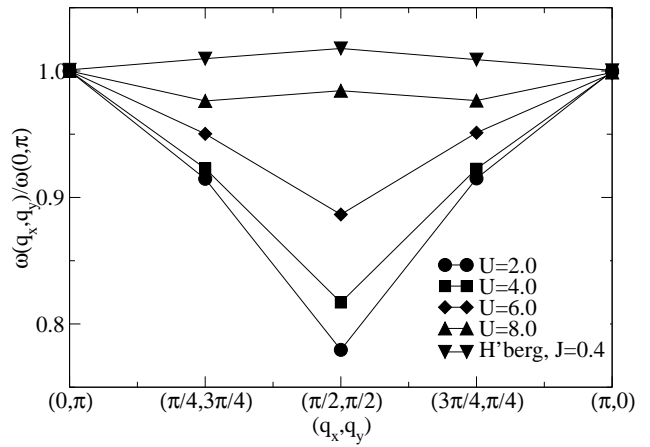}
\vskip1mm
\caption{Magnon dispersion along the magnetic zone boundary for different values
of $U$ for a $N$=8x8 lattice at $\beta=3.0$. The Heisenberg plot corresponds to
$U=10.0$. Error bars are of the order of symbol sizes.} 
\label{fig:fig3}
\end{figure}

We now discuss the
variation of the dispersion along the zone boundary in the 2D Hubbard model
as a function of the on-site interaction parameter, $U$. Figure \ref{fig:fig3} shows the
results of our simulation for a $N$=8x8 lattice at $\beta=3.0$. We have focused
primarily on the deviation of the magnon energy at the center of the zone
boundary relative to the edges. To that end, we have scaled the magnon energies
by their value at $(0,\pi)$. The deviation is seen to be maximum for the
smallest value of $U$ studied and decreases with increasing $U$ -- becoming essentially
flat within statistical error at $U=8.0$. Unfortunately, for $U > 8$, the results
of our simulation for the 2D Hubbard model become too noisy. However, in this
limit of large $U$, the Hubbard model maps on to the Heisenberg model. Hence the 
dispersion for larger $U$ values is expected to be be qualitatively similar to that 
for the Heisenberg model. We have also shown in the figure
the dispersion obtained for a Heisenberg model
with exchange parameter $J$ corresponding to $U=10.0$. 
The point $U=8.0$ marks a transition to 
``Heisenberg-like'' behavior. This value of the transition point is consistent with
that found from specific heat measurements of the 2D Hubbard model \cite{spheat}.

To investigate other possible sources for the nature of magnon energy dispersion
along the magnetic zone boundary, we have studied the effects of spin-phonon
coupling within the framework of the 2D Heisenberg model. The model involves 
coupling of the spins to dynamic optical bond phonons and is given by the Hamiltonian

\begin{equation}
H = -J\sum_{\langle{\bf i,j}\rangle}(1 + \alpha(a_b^{\dagger}+a_b)){\bf S_i. S_j}
+ \omega_0\sum_ba_b^{\dagger}a_b
\end{equation}

\noindent where $b$ denotes the bond between the lattice sites {\bf i} and {\bf j}
and the operator $a_b^{\dagger}(a_b)$ creates(annihilates) a phonon on bond $b$.
$\omega_0$ represents the bare phonon frequency and $\alpha$ measures the strength
of the spin-phonon interaction.

\begin{figure}
\centering
\epsfxsize=8.3cm
\leavevmode
\epsffile{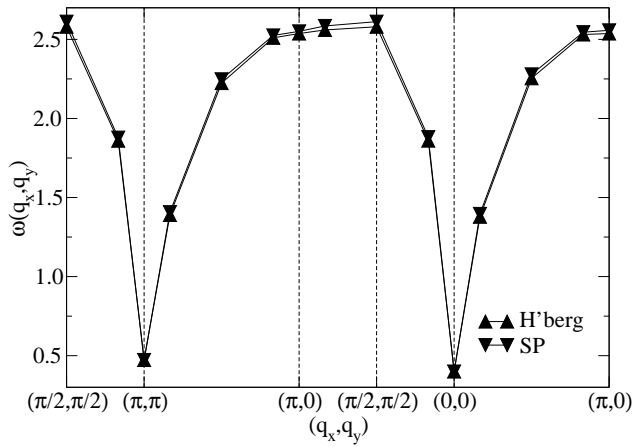}
\vskip1mm
\caption{Magnon dispersion along the same path as in Figure \ref{fig:fig2} for a $N$=8x8 lattice at $\beta=5.0$ for the pure Heisenberg model and the spin-phonon model with $\alpha=0.1$
and $\omega_0=0.25$.} 
\label{fig:fig4}
\end{figure}

The spin-phonon  model was studied using the SSE QMC method. Our results indicate 
that over any reasonable range of spin-phonon coupling strength and bare phonon
frequency, the magnon dispersion along the magnetic zone
boundary is qualitatively the same as the pure Heisenberg model. 
Figure \ref{fig:fig4} shows the variation of the magnon energy along the same
path in momentum space as considered in Figure \ref{fig:fig2} for the pure Heisenberg
model and the spin-phonon model with $\alpha=0.1$ and $\omega_0=0.25$
for a lattice of size $N$=8x8 at $\beta$=5.0. The value of $\beta$ is
chosen such that $\beta J$ is the same as $\beta t$ for the data presented
in Figure \ref{fig:fig2} with $J=4t^2/U$. This allows for a direct comparison between the
two sets of data. Once again we found that while the data near $(\pi,\pi)$
and (0,0) are strongly temperature dependent, the magnon energies along 
the magnetic zone boundary from $(\pi,0)$ to $(\pi/2,\pi/2)$ have converged
close to their ground state values. The data suggests
that the deviation in the magnon dispersion observed for
$\mbox{La}_2\mbox{CuO}_4$ from that found in the Heisenberg model cannot be explained
by the type of spin-phonon coupling considered here. It remains to be seen whether
a more realistic treatment of the electron-phonon coupling 
in the cuprates can account for this behavior.

To summarize, we have used the det QMC to show that the experimentally observed magnon dispersion 
along the magnetic zone boundary in $\mbox{La}_2\mbox{CuO}_4$ can be
reproduced by the  2D Hubbard model, using reasonable $U/t$ values.
The deviation of the magnon energy at the center of the zone boundary is
maximum for the smallest value of $U$ considered, where the charge fluctuations are the 
largest, and decreases with increasing $U$. For $U\approx 8t$, the dispersion is 
flat within statistical errors. The dispersion for larger values of $U$ is 
qualitatively similar to that of the pure Heisenberg model. Thus, at $U\approx 8t$,
there is a transition to ``Heisenberg-like'' behavior where the effects of the
charge fluctuations become negligible. 
An interesting question is whether this intermediate $U/t$ regime of the Hubbard model is
equivalent to a Heisenberg plus ring exchange term. At really small $U/t$
the electrons will be strongly delocalized, so that a description in terms
of a spin Hamiltonian would break down. However, at the intermediate $U/t$,
found in our calculations, a higher order perturbative treatment maybe
justified leading to a Heisenberg plus ring exchange Hamiltonian.

We would like to thank Anders Sandvik for many useful discussions.
This work was supported in part by NSF grant number 9986948. The
numerical simulations were carried out in part at the Condor Flocks at
University of Wisconsin, Madison and NCSA, Urbana, Illinois.

\null

\end{document}